# Stimulation technology for brain and nerves, now and future

Masaru Kuwabara[1,*], Ryota Kanai[1,*],

1 Araya, Inc., Chiyoda, Tokyo, Japan

* m.kuwabara@araya.org, * kanair@araya.org

**Abstract**

In individuals afflicted with conditions such as paralysis, the implementation of Brain-Computer-Interface (BCI) has begun to significantly impact their quality of life. Furthermore, even in healthy individuals, the anticipated advantages of brain-to-brain communication and brain-to-computer interaction hold considerable promise for the future. This is attributed to the liberation from bodily constraints and the transcendence of existing limitations inherent in contemporary brain-to-brain communication methods. To actualize a comprehensive BCI, the establishment of bidirectional communication between the brain and the external environment is imperative. While neural input technology spans diverse disciplines and is currently advancing rapidly, a notable absence exists in the form of review papers summarizing the technology from the standpoint of the latest or potential input methods. The challenges encountered encompass the requisite for bidirectional communication to achieve a holistic BCI, as well as obstacles related to information volume, precision, and invasiveness. The review section comprehensively addresses both invasive and non-invasive techniques, incorporating nanotech/micro-device technology and the integration of Artificial Intelligence (AI) in brain stimulation.

**Introduction**

In individuals with conditions such as paralysis due to ALS, cervical spine injuries, or quadriplegia, direct information exchange between the brain and a computer yields a profound impact on their quality of life. Even in healthy individuals, brain-to-brain communication and brain-to-computer interaction are expected to bring tremendous benefits in the future, as it frees them from their bodies and transcends the current limitations of brain-to-brain communication.

Neurotechnology has experienced a growing application in various industries in recent years. This field, involving the direct transfer of information to and from the brain and nervous system, is anticipated to revolutionize education, health, communication, and medical applications, including treatments and prosthetics. Despite significant progress in reading brain activity, the development of technology for "writing in" is still in its nascent stages. To achieve a complete Brain-Computer-Interface (BCI), the establishment of bidirectional communication between the brain and the external world is essential.

Existing methods face challenges related to the precision, and invasiveness. Moreover, while neural input technology spans a diverse range of methods and is rapidly evolving, there is a lack of comprehensive review papers summarizing the latest input methods. This review aims to concentrate on emerging brain input technologies.

The considerable benefits of brain stimulation include:

- Facilitating the transfer of linguistic information, particularly crucial in situations where individuals cannot rely on their sight or hearing.
- Enabling the transmission of concepts that defy effective communication through language, such as non-declarative memories and abstract concepts.
- Contributing to brain research as a tool, especially in demonstrating causality by establishing a direct link between input and resulting phenomena.
- Holding the potential to significantly enhance cognitive function and memory.
- Offering avenues for the alleviation and treatment of neurological and psychiatric conditions.

This review focuses on the first three benefits listed above.

Sections cover both invasive and non-invasive techniques, including nanotech/micro-device technology and the integration of AI in brain stimulation. In the invasive technology realm, we cover optogenetics, intracortical microstimulation (ICMS), direct electrical stimulation (DES), and cochlear/retinal prostheses. Non-invasive technology will be explored, with a specific focus on focused ultrasound (FUS).

## 1.1 Optogenetics

To achieve complete manipulation of the brain, manipulation at the single-neuron level becomes necessary. Optogenetics emerges as a potential solution, offering the capability to stimulate neurons with cell-type specificity (i.e., pyramidal cells, parvalbumin cells, somatostatin cells). This section aims to introduce the latest advancements in optogenetics, primarily focusing on its applications in research while also considering its potential clinical applications.

Optogenetics came to prominence in 2005 following the work of Ed Boyden and Karl Deisseroth (Boyden et al., 2005). Optogenetics is a technology that allows selective expression of opsin in targeted cells and light stimulation of opsin-expressing cells specifically. Opsins are a group of proteins made light-sensitive via the chromophore retinal. Excitatory opsins can be utilized to stimulate activity within neurons whereas inhibitory opsins can suppress activity. The technique of cell-selective stimulation of neurons on a microscopic scale is unique to optogenetics, offering millisecond temporal precision.

We first present examples of using optogenetics to control or transmit information to an animal. Optogenetics has been successfully used to control the limbs of rodents (Soma et al., 2019; Watanabe et al., 2020). As brain-to-brain interfaces, Lu et al. succeeded to connect two animal's brains and synchronized two animals' locomotion behavior (Lu et al., 2020). In primate studies, limb (Ebina et al., 2019) and saccade (Inoue et al., 2015) were controlled by optogenetics. In Mongolian gerbils, it was found that optogenetic stimulation was spatially more selective than monopolar electrical stimulation at all activation strengths and outperformed bipolar electrical stimulation (Dieter et al., 2020). In their paper, increased spectral selectivity of optogenetic stimulation over electric auditory nerve stimulation suggests that cochlear optogenetics can, indeed, increase the frequency resolution of artificial sound encoding.

In humans, optogenetics has not yet been FDA-approved or clinically studied in the central nervous system, but there are attempts to apply optogenetics to humans (White et al., 2020). Clinical trials have been conducted in the retina and cochlea. The fact that optogenetics performed better than electrical stimulation in Mongolian gerbils suggests the possibility of practical application. We will discuss this further in section 1.4.

There have been technological advancements in light delivery. Wu et al. reported mechanoluminescent nanoparticles that act as local light sources in the brain when triggered by brain-penetrant focused ultrasound (FUS) through the intact scalp and skull, a concept termed sono-optogenetics.

Mechanoluminescent nanoparticles can be delivered into the blood circulation via intravenous injection, recharged by 400-nm photoexcitation light in superficial blood vessels during circulation, and turned on by FUS to emit 470-nm light repetitively in the intact brain for optogenetic stimulation (Wu et al., 2019). Chen et al. used upconversion nanoparticles (UCNPs) to employ similar method. UCNPs emit a spectrum peaking at 450 nm and 475 nm upon excitation at 980 nm near-infrared light, which possesses the ability to penetrate tissue (Chen et al., 2018).

Optogenetics can stimulate single cell or specified multiple cells in vivo with 2-photon holographic patterned excitation. This allows for much more precise optogenetic manipulation than previously possible, e.g., by sequentially removing unnecessary neurons to identify the minimal subset of neurons that can alter behavior (Adesnik and Abdeladim, 2021; Chen et al., 2019; Packer et al., 2015). Wireless control of a small device equipped with micro-LED has been reported, which is crucial for chronic implants. Together with nanotechnology, optogenetics has the potential to bring BCI to a different level (Balasubramaniam et al., 2018; Goncalves et al., 2017). Many types of opsins have been identified, each with distinct peak activation wavelengths, decay kinetics, and varying durations of the opening channel (Carrasco-López et al., 2020; Yizhar et al., 2011). Furthermore, a recent study reported a new variant of channelrhodopsin that responds to red light with an extremely large photocurrent, allowing it to penetrate much deeper into brain tissue compared to blue light. This breakthrough enabled optogenetic modulation of the deep brain region up to a depth of 7 mm. This capability was demonstrated by targeting the brain stem of a mouse without the need for intracranial surgery (R. Chen et al., 2021).

Optogenetics has the following advantages. 1) Bidirectional control of cellular activities is feasible. 2) Specific cell-type targeting is achievable by controlling the gene delivery process to express light sensitive proteins predominantly in the cell-type of interest. 3) Independent control of neurons in the same area is possible through the combinational usage of multiple types of opsins and multi-frequency wavelength light. 4) The inherent parallel nature of optics can aid in manipulating neural activities in large-scale neural networks, particularly in the cortex.

On the other hand, optogenetics has the following shortcomings. 1) It requires infecting targeted cells. Additionally, opsin expression can take up to six to eight weeks from injection to reach sufficient level. 2) Light transmission falls off quickly through neural tissue, being reduced by 50% after 100 μm and by 90% after 1 mm when using 473 nm blue light. Therefore, the light should be emitted from the close proximity to the brain tissues. 3) The long-term effect on neurons of retrovirus insertion into the genome, expressing non-human opsin proteins, and the impact of prolonged exposure to light are unknown. Gene delivery methods must prove to be safe and stable.

Optogenetics has contributed to neuroscience research in the last two decades and has the potential to be applied in humans with high expectations.

**1.2 Intracortical microstimulation (ICMS)**

Optogenetics has been a robust research method and still requires time to evolve into an application-level tool in humans. In this section, we introduce intracortical microstimulation (ICMS) as a potent method that is already in use for human applications with a long history.

The most traditional and well researched method to stimulate central nervous system is electrical stimulation. ICMS is employed to map neural circuits and restore lost sensory modalities, such as vision, hearing, and somatosensation. To transfer information at a high rate with precision, the optimization of the stimulation pattern and channel alignment for stimulating neurons is necessary.

Physiologically, ICMS is thought to activate both inhibitory and excitatory populations of cells (Butovas and Schwarz, 2003), and it is not believed to evoke natural patterns of cortical activity since these patterns differentially propagate to downstream structures (Millard et al., 2015). ICMS with intensities ranging from 2 to 150 mA can activate neurons within 50–500 μm of a stimulating electrode, corresponding to 60–62,000 neurons in monkey V1 (Tehovnik and Slocum, 2007). At maximum amplitudes, it could reach 2 mm. The modality of sensation is determined based on the stimulation area. Here, we introduce applications in three modalities, i.e., somatosensation, vision and audition.

With ICMS, the most well-researched modality is somatosensation. A Multielectrode Array (MEA) in the primary somatosensory area could achieve a resolution of a few millimeters in finger/palm sensation (Kramer et al., 2019). Somatosensory feedback is essential for efficient and precise movement and to manipulate objects effectively. Sensory and motor functions do not exist as distinct processes. Complex motor plans and desired outcomes, along with sensory feedback, are integrated and compared in our brain to make appropriate adjustments. This process is referred to as sensory-motor coupling, and it becomes particularly crucial for tasks that demand fine dexterity (Bensmaia and Miller, 2014; C. Hughes et al., 2020; Shokur et al., 2021). From this perspective, converting somatosensory feedback into other modalities, such as language, sound, or visual cues, is challenging and requires a greater cognitive load, resulting in longer times to support motor functions.

Somatosensory stimulation typically induces unnatural sensations, such as tingling or buzzing. However, Armenta Salas et al. found reliable elicitation of natural cutaneous and proprioceptive sensations by adjusting stimulus amplitudes and frequencies, which were weaker than previously utilized parameters (Armenta Salas et al., 2018). Thus, there is still significant room for improvement in evoking truly natural and biomimetic sensations via ICMS (C. Hughes et al.,

2020).

The prospect of providing artificial vision for the blind has led to significant interest in developing visual cortical prostheses, with several ongoing and planned clinical trials. A phosphene, used in visual prosthetics, is the phenomenon of perceiving light without actual light entering the eye, induced by electrical stimulation of the visual streams. Monkeys immediately recognized them as simple shapes, motions, or letters with 1024-channel prosthesis in areas V1 and V4 of the visual cortex (Chen et al., 2020). In human, MEA (96ch, each channel 400μm apart) in primary visual cortex conveyed simple lines, shapes, some letters, and object boundaries (Fernández et al., 2021). Fernández et al. found that 4-6 sessions of 10-15 minutes of training enhanced performance, and the effect persisted for months.

In auditory cortex stimulation, animal experiments have demonstrated that animals can discriminate the frequency of ICMS (Otto et al., 2005) and differentiate temporal and spatial patterns (Scheich and Breindl, 2002) in rats and Mongolian gerbils, respectively. A decoding study of speech from macaques' auditory cortex suggested that decoding approaches may further be useful in informing future encoding strategies (Heelan et al., 2019). However, human studies of ICMS in primary auditory cortex (e.g., (Dobelle et al., 1973)) are limited. This limitation arises due to the reduced accessibility of the primary auditory cortex compared to other primary cortical areas. Moreover, the primary auditory cortex exhibits less distinct tonotopy due to its more plastic nature, resulting in a more complex coding of perceptual sound features. ICMS applied to the auditory cortex has been shown to elicit complex and abstract auditory percepts (Dobelle et al., 1973; Histed et al., 2013; Lim et al., 2011). Another reason may be that, since targeting the cochlea, auditory nerve, or auditory midbrain has already shown its potential, especially with the huge success of cochlear implants, the necessity of ICMS in the auditory cortex is limited unless damage occurs to these organs.

Although subjects can typically detect microstimulation of the primary sensory, visual, auditory, and motor cortex, they are generally unable to detect stimulation of most parts of the cortex without extensive practice. However, with practice, detection of stimulation in any part of the cortex becomes possible. The fact that the adult brain retains enough plasticity to learn to detect electrical activation anywhere in the cortex has implications for the development of cortical sensory prosthetics (Histed et al., 2013).

Stimulation parameters have been investigated, and their exploration may contribute to optimizing stimulation efficiency. For instance, Hughes et al. discovered that increasing the frequency led to more intense percepts on certain electrodes but resulted in less intense percepts on others (Hughes et al., 2021). The response to the change in stimulation frequency differs depending on the area. Utilizing spatiotemporal patterns of intracortical microstimulation, Balasubramanian et al. found that reaction time increases significantly when stimulation is delivered against, but not with, the natural propagation direction in the primary motor cortex (Balasubramanian et al., 2020). There are also studies to optimize the stimulation pattern with deep learning (Küçükoğlu et al., 2022).

Long-term stability of electrodes is also an important aspect of human ICMS use. Hughes et al. observed iridium oxide electrodes for 1500 days, providing evidence that ICMS in the human somatosensory cortex can be maintained over extended periods without deleterious effects on recording or stimulation capabilities (C. L. Hughes et al., 2020). Rosenfeld et al. investigated the effects of long-term stimulation in the sheep cortex. They achieved over 2,700 hours of stimulation without observing any adverse health effects (Rosenfeld et al., 2020).

In summary, as ICMS stimulate the neural population of the vicinity, the interaction with the brain is mostly topography dependent. Since it penetrates cortex, it is highly invasive, and covering broad area with this method is challenging (Nurmikko, 2020).

**1.3 Direct electrical stimulation (DES)**

Theoretically, ICMS can cover the entire brain, but implementing this across the entire human brain is currently considered too invasive and impractical. DES is characterized by reduced invasiveness, as electrodes are positioned on the brain's surface without penetrating the cortex. This configuration allows DES to encompass a wide area of the human brain cortex. This section focuses on the types of information that can currently be represented, their resolution, and emphasizes applied research in humans.

Direct electrical stimulation (DES) [or when applied particularly to cortex, known as direct cortical stimulation (DCS), or direct electrical cortical stimulation (DECS)] is also known as cortical stimulation mapping for functional mapping of the cortex, which aids surgery by identifying brain areas that are critical for important functions. DES experiments have been also performed as part of monitoring studies for intractable epilepsy (Borchers et al., 2012). The application of electrical stimulation mapping of the brain for clinical use has been in practice for approximately a century (Ritaccio et al., 2018). The cortico-cortical evoked potential (CCEP) technique was originally introduced as an extra-operative procedure preceding epilepsy surgery to preserve the language center and arch bundle by tracking various brain networks (Matsumoto et al., 2004), which is caused by the electrophysiological propagation of the response through white matter bundles, from the stimulated cortical region to the distant recording site by DES. DES is less invasive method compared to ICMS as the electrodes do not penetrate cortex.

For the implementation of DES, it is common to mainly use ECoG or DBS electrodes. ECoG possesses traits that fall between MEA and EEG in terms of invasiveness and signal quality. In comparison to ICMS, DES of the human cortex using larger electrodes injects current over a broader surface area. Consequently, the application of large amounts of current could lead to greater activation with the potential to spread to a larger area. On average, the amount of charge per pulse delivered in DES is 280 times greater than that in ICMS (0.2–16 μC for DES and 0.4–37.5 nC for ICMS). Given that the electrode surfaces for the two techniques are very different, the charge densities are likely to be in much the same range (Vincent et al., 2016).

ECoG can be inserted epidurally, subdurally, or on the cortex, with a trade-off in infection risk and signal quality. The electrodes include 1.5 mm diameter circular contacts with 4 mm spacing for "micro"-ECoG electrodes, and 2.3–3 mm diameter contacts with 10 mm spacing for "macro"-ECoG electrodes (Caldwell et al., 2019). Kramer et al. compared MEA, standard ECoG and micro-ECoG (with 1.2 mm diameter contacts with 3 mm electrodes) and found that micro-ECoG represents a promising combination of minimal invasiveness and improved signal quality. It provides an excellent balance between spatial cortical coverage of the hand area in the primary somatosensory cortex and high-density resolution (Kramer et al., 2019). Ho et al. reported high density micro-ECoG with 2,116 channels covering a surface area of 1.92 cm$^2$ for both recording and stimulation, demonstrating swift insertion capabilities (Ho et al., 2022).

Similar to ICMS, DES has been used for sensory feedback in BCI applications (C. Hughes et al., 2020). Cronin et al. demonstrated that subjects utilized DES in the sensory area and, it enhanced motor task performance (Cronin et al., 2016).

Oswalt et al. reported that one blind subject correctly identified four different letter shapes ("W", "N", "M", "U") at 93% accuracy with DES in V1-V3, without prior training (Oswalt et al., 2021). Another research group used a dynamic sequence of phosphenes to trace forms, which were presented and recognized rapidly by blind participants, up to 86 forms per minute (Beauchamp et al., 2020). Armenta Salas et al. observed a significant impact on the perceived brightness of phosphene in relation to the stimulation sequence. Moreover, it appears that this stimulation order interacts with the current amplitude level (Armenta Salas et al., 2022).

At present, the reported studies focus on developments using either surface/subdural electrodes or intracortical electrodes. Notably, no projects have proposed a hybrid approach (Pio-Lopez et al., 2021). For improved resolution and coverage in visual and sensory prostheses, this combination may produce a similar effect by replicating the distinction between the fovea and periphery in the retina, or the differentiation between fingers and broader body parts in the sensory area.

**1.4 Cochlear and retinal prosthesis**

The previously presented examples primarily exist at the research level, encompassing both basic and applied research involving direct brain stimulation. This section now aims to introduce technologies deployed at a practical and clinical level. The focus shifts towards instances that have transitioned into commercialization and are presently employed as neuroprosthetics, albeit operating at the peripheral nerve level.

*Cochlear prosthesis*

Currently, the cochlear implant is considered the most successful neuro-prosthesis, as it typically enables open speech comprehension in quiet environments. According to the World Health Organization's 2010 census, approximately 360 million people worldwide have hearing impairments (Sun et al., 2019). Cochlear implants have facilitated speech understanding for the majority of the approximately 700,000 users worldwide (Keppeler et al., 2021). However, there are currently differences in effectiveness among patients. While, in many cases, it is possible to achieve a level of hearing that allows for speech recognition, individuals may encounter challenges in communicating in noisy environments or with tonal languages. Additionally, there is a limited ability to appreciate music and accurately understand everyday sounds, such as the sound of pouring water (Dornhoffer et al., 2021).

Since 1957, both devices and algorithms have undergone improvement. While the number of stimulation channels has increased over time, current understanding suggests that performance levels plateau at approximately 22 channels (Cucis et al., 2018). The limitation arises because the area affected by electrical stimulation is not narrow enough in the tonotopicity of the cochlea. Even if the channels were denser, achieving high resolution in the specific stimulation of each frequency area in the cochlea remains a challenge. Consequently, the development of cochlear implants employing optogenetics has garnered significant attention (Weiss et al., 2016). In gerbils, optogenetics has already been shown to induce neural activity more similar to that of natural sound than electrical stimulation as optical cochlear stimulation promises lower spatial spread of neural excitation and spectral selectivity as mentioned in section 1.1 (Dieter et al., 2019; Keppeler et al., 2021). Despite the need to address energy requirements, temporal bandwidth, frequency, and intensity resolution in coding to integrate optogenetics into cochlear implants (Moser, 2015), several groups are presently working towards the first-in-human application of optogenetics in the human brain. The first human trials are likely to occur within the next decade (White et al., 2020). Another potential method to achieve high spectral selectivity is near-infrared stimulation, which does not necessitate AAV infection. Nevertheless, there is a risk of heat accumulation in the tissue at high pulse repetition rates. Therefore, researchers are exploring optimal pulse shapes and combined optical/electrical stimulation to mitigate this issue (Littlefield and Richter, 2021; Richardson et al., 2020).

The coding strategy, which converts speech into electrical signals, has been improved many times, contributing to increased accuracy (Sun et al., 2019). Alongside advancements in device technology, enhancements in coding strategies facilitated by machine learning and/or other algorithms have contributed to the improvement of cochlear implant technology. This area is also identified as one of the most promising research avenues in the near future (Crowson et al., 2020; Kang et al., 2021).

*Retinal prosthesis*
The most important sense is sight. Currently, millions of people worldwide are facing the challenges associated with severe vision loss due to increasing longevity. Visual prosthesis through brain stimulation has been previously discussed, and now we will focus on retinal prosthesis. There are two major types of prostheses: one is based on electrical stimulation, and the other is based on optogenetics. First, we introduce state-of-the-art technology based on electrical stimulation.

Distinguished by surgical approach, there are four types of electrical retinal prostheses; epiretinal prostheses (in which electrodes are placed on the retina), subretinal prostheses (in which electrodes are placed beneath the retina), suprachoroidal prostheses (in which electrodes are placed in the suprachoroidal space), and intrascleral prostheses (where the electrodes are placed within a pocket in the sclera)(Ayton et al., 2020).

Over five hundred patients have been implanted retinal prostheses globally over the past 15 years (Ayton et al., 2020). There are several prosthetic devices with a broad range of channel numbers, ranging from 16 to 1500 channels (EyeWiki® https://eyewiki.aao.org/Retina_Prosthesis). In clinical trials, these devices have provided visual acuities as high as 20/ 460(0.0435) in the best case, enabled coarse navigation, and even allowed for reading short words, although sometimes requiring prolonged scanning to identify letters (Abbasi and Rizzo, 2021; Palanker et al., 2020).

For retinitis pigmentosa patients, AAV-mediated optogenetic vision restoration has been approved, as indicated by clinical trials NCT02556736 and NCT03326336 (Garita-Hernandez et al, 2018). To date, four clinical trials involving optogenetics have been conducted, all of which were for the treatment of Retinitis Pigmentosa (White and Whittaker, 2022). In clinical trials,

Sahel et al. presented the first evidence that the injection of an optogenetic sensor-expressing gene therapy vector combined with the wearing of light-stimulating goggles can partially restore visual function in a patient with pigmentosa patients who had a visual acuity of only light perception (Lindner et al., 2022). The results of clinical trials "RESTORE" (NCT04945772, NCT04919473) have not been published yet but there is a press report which states that "Some of the patients even gained the ability to read letters on a wall or even the large text in a newspaper, use a cell phone, watch television, and could even thread a needle.", indicating that it could be a new promising intervention (Ophthalmology TIMES; https://www.ophthalmologytimes.com/view/optogenetic-gene-therapy-restores-vision-in-11-rp-patients). The need to develop an implant capable of safely introducing light within the body means that optogenetic trials for other conditions may take another 5–10 years (White and Whittaker, 2022).

*Summary*

Prostheses using electrical stimulation for both the cochlea and retina are already available. In particular, the cochlea prosthesis has reached a very high practical level after years of development, although there are still some problems in terms of sound quality and individual differences in outcomes. The key challenge in the context of retinal prosthesis lies in the difficulty of augmenting spatial resolution. Despite the 378-channel system demonstrating a current best acuity of 20/460, it falls short of the 20/400 (0.05) threshold for blindness. In various clinical trials, the acuity has not reached even 20/2000 (0.01). Further innovations in optogenetics are also anticipated.

In the case of both electrical stimulation and optogenetics, temporal resolution for information transfer can be achieved satisfactorily, but achieving high spatial resolution is very challenging. As evidenced by the fact that the visuospatial sketchpad holds less than 0.5 s input in Buddley's model, while the phonological loop holds 2-4 s input, audition is innately a modality that is superior in extracting information from time series, and that could be related to the success of the cochlear prosthesis. When considering the information transfer rate (ITR), cochlear prostheses currently exhibit the highest rate by far. From the perspective of information input to the brain via BCI, the peripheral nervous system will be the initial target. Once that becomes sufficient, attention will shift to the central nervous system, opening up further possibilities as the next target.

In their paper, Abbasi and Rizzo states "We hypothesize that further advances in engineering will be insufficient to deliver the hoped-for level of vision; rather, we believe that a deeper and more nuanced understanding of neuroscience, especially the neural code of vision, is now the limiting factor for success." (Abbasi and Rizzo, 2021). A compensatory approach that could enhance perceptual quality involves selective, or at least significantly biased, stimulation of either ON- or OFF-pathways. This has been achieved in vitro and could also be achieved through improved computational encoding to mimic retinal signal processing.

## 2.1 Transcranial focused ultrasound (FUS)

Up to this point, we have outlined invasive methodologies. Nevertheless, the aspect of noninvasiveness holds paramount significance in the context of brain stimulation. Until firm assurances regarding both usefulness and safety are established, it is imperative that brain input techniques adopt a less invasive approach to become a widely accepted technique, even among healthy subjects. While techniques like transcranial direct current stimulation (tDCS), transcranial alternating current stimulation (tACS), and transcranial magnetic stimulation (TMS) have long been recognized as noninvasive forms of brain stimulation, their spatial resolution, exceeding a few centimeters, remains limited. FUS is an exceptional method that addresses this issue. In this section, we aim to introduce several examples of its use, primarily in humans, along with some research cases.

Transcranial focused ultrasound (FUS) is emerging as a novel neuromodulation approach, combining noninvasiveness with a sharp focus, even in regions deep within the brain, such as the human thalamus at a depth of 6 cm (Legon et al., 2018). Ultrasound can penetrate biological tissues, including the skull, concentrating its energy into a small targeted area of the brain with a 4 mm resolution in the human cortex (Legon et al., 2014).

The MRI-guided FUS procedure enables the precise ablation of targeted brain areas. Since its introduction, there has been a surge in the clinical use of high-intensity FUS, exemplified by the FDA approval of FUS for treating essential tremor and Parkinson's disease. Additionally, in other countries, FUS has gained approval for treating depression, neuropathic pain, and obsessive-compulsive disorder ("Focus Feature: State of the Field Report 2023," n.d.; Meng et al., 2021).

In this review, we focus on low-intensity FUS, which modulates neural activity without causing damage through cavitation or heating. Given its novelty, many researchers and clinicians may not be fully aware of the potential applications and promises of this technique. We aim to introduce its mechanism, clinical potential, and its utility as a research tool.

As mechanisms of FUS, it is known that acoustic radiation force alters membrane conformational state and permeability of mechanosensitive ion channels, such as Piezo1, TRPP1/2 complex, TRPA1 and TRPC1 mechanoreceptors, as well as voltage-gated calcium, sodium, and potassium channels in neuronal membranes. This phenomenon is referred to as electrophysiological-mechanical coupling. Other known mechanisms of FUS include microtubule resonance, slight cavitation, thermal mechanisms, and sonoporation (Dell'Italia et al., 2022; Jerusalem et al., 2019; Yoo et al., 2020).

From clinical perspective, it is shown that FUS can safely, reversibly, and repeatedly open blood-brain barrier in humans (Lipsman et al., 2018). A clinical trial has indicated FUS is a promising alternative form of neuromodulation for drug-resistant epilepsy (Bubrick et al., 2022). Additionally, FUS alleviates pain sensitivity (Badran et al., 2020), improve mood in depressive patients through neuromodulation (Hameroff et al., 2013; Sanguinetti et al., 2020), and exhibits potential for neurorehabilitation, with ongoing clinical trials (W. Lee et al., 2021).

Next, we introduce some works that demonstrate the potential of FUS as a research tool and information transfer method. Thus far, we have discussed cases where FUS targeted a single point. However, continuous two-dimensional ultrasound holograms enable the efficient generation of uniform and accurate continuous ultrasound patterns at a few millimeter scale, as demonstrated in studies such as alphabet-shaped stimulation on the retina (Naor et al., 2012). Designed a dual-crossed transducer system that achieved high spatial resolution, and the full-width at half maximum focal volume of our dual-crossed transducer system was under 0.52 mm$^3$ in the mouse brain (S. Kim et al., 2021). Kim et al. developed a wireless wearable ultrasound neurostimulator designed for small animals, such as mice, enabling stimulation in freely moving animals (E. Kim et al., 2021). As behavioral effect of FUS, for example, it has been reported that FUS can affect visuomotor behavior in monkeys (Deffieux et al., 2013).

Stimulus parameters, such as intensity and stimulus pattern (duty cycle), were theoretically shown to specifically stimulate cell type-selective network regulation either excitatory or inhibitory (Plaksin et al., 2016). Additionally, empirical studies have indicated that pulse repetition frequency, duty cycle, sonication duration, and tissue properties may influence the inhibition or excitation of cortical neurons (Kubanek, 2018; Zhang et al., 2021).

In human, phosphene (Lee et al., 2016b) and tactile limb sensations (Lee et al., 2016a) can be elicited by FUS. The first demonstration of FUS-based brain stimulation in the context of brain-to-brain interface achieved a success rate of 96.5 ± 3.1% (mean ± s.d.) (Lee et al., 2017). Cain et al. showed that FUS on subcortical area had both real-time and delayed effects (Cain et al., 2021). Legon et al. targeted human thalamus and found both behavioral and neural activity changes (Legon et al., 2014). The FDA guidelines for cephalic ultrasound suggest a maximum safety value range of $I_{spta}$ < 94 mW/cm$^2$, $I_{sppa}$ < 190 W/cm$^2$, and MI < 1.9 to avoid cavitation and heating in humans (Fomenko et al., 2020).

Altogether, FUS is a promising new non-invasive method with significant potential to alter short-term brain excitability and connectivity, induce long-term plasticity, and modulate behavior. Recent studies also provide initial evidence regarding its safety and feasibility in humans (Arulpragasam et al., 2022). The weakness of FUS is that the stimulus intensity is weak to avoid heat and cavitation in targeted tissue. Currently, only about half of the subjects can induce phosphine, and no motor function has been induced in humans, and the stimulus parameters may need to be improved. Notably, adverse effects were found in 3.4% of subjects (Sarica et al., 2022). While FUS is anticipated to become operational in humans relatively soon, further research is needed regarding writing information and its clinical application.

## 3.1 Nanotech/ micro-device

Ultimately, the ideal method is to stimulate the desired target area minimally invasively at the single-cell level. Although the technology is still in its infancy for human use, there are research areas with the potential to achieve this. While deep brain stimulation (DBS) is expensive, costing approximately $200,000, the cost of remedies could be reduced if nanotechnology becomes more widely used. Nanomaterials and/or microdevices for neuromodulation have the potential to be minimally invasive, cost-effective, and biocompatible compared to conventional tethered devices (Knoben et al., 2022). In this section, we present the state-of-the-art in nanotechnology and microdevices, which are rapidly developing in recent times.

Nanomaterials are defined as low-dimensional materials with building units smaller than 100 nm, at least in one dimension. (Biswas and Wu, 2005). Nanotransducers have the capability to wirelessly transduce external electric fields, light, magnetic fields, or ultrasound waves into a local signal (light, thermal, mechanical, electrical, or chemical) within the region of interest (Li et al., 2021). Converted local mechanical force can modulate membrane capacitance or activate mechano-sensitive channels, such as transient receptor potential cation channel subfamily V member 4 (TRPV4) and piezo-type mechanosensitive ion channel component (PIEZO) channels. Local temperature changes can modulate neuron activity by changing membrane capacitance or activating temperature-gated ion channels, such as the transient receptor potential cation channel subfamily V member 1 (TRPV1), the transient receptor potential cation channel ankyrin 1 (TRPA1), and the temperature-gated chloride channel anoctamin 1 (TMEM16A). Light can stimulate the rhodopsin channel, as introduced in the optogenetics section. For instance, gold nanorods have the capability to convert infrared light into heat, thereby stimulating temperature-sensitive ion channels (Nelidova et al., 2020). $CoFe_2O_4$-$BaTiO_3$ magnetoelectric nanoparticles (MENs) can convert magnetic field into electrical signal (Nguyen et al., 2021). Noninvasive neuromodulation is accomplished by using ultrasound to uncage neuromodulatory drugs from nanoparticle drug carriers (Wang et al., 2018). By applying antibody onto the surface of nanoparticle, the target location of neurons stimulated can be precisely determined (Liu et al., 2022). Light, magnetic fields and ultrasound can penetrate skull and brain tissue invasively, specifying the stimulating area by focusing on nanotransducers and/or power sources.

In recent years, microdevices have rapidly become smaller, wireless, multi-channel, and tissue-friendly, incorporating features such as flexibility (Y. Chen et al., 2021; Hong et al., 2021; Vázquez-Guardado et al., 2020). For instance, a flexible and wireless microelectrode array designed for surface recordings and stimulation as part of a visual prosthesis boasts over 65,000 electrodes (Garcia-Etxarri and Yuste, 2021). Piezo and other microdevices have wireless power supplies via ultrasound, which enables wireless uLEDs and electrical stimulation. Ultrasound serves as an alternative for implementing wireless systems without relying on electromagnetic waves. An exemplary work in this domain is StimDust, a pioneering system for wireless neural stimulation that employs 1.7 mm$^3$ leadless and wireless StimDust stimulation motes (Piech et al., 2020). J. Lee et al. reported wirelessly networked and powered electronic microchips known as neurograins. These microchips, smaller than 0.1 mm$^3$, can autonomously perform neural sensing and electrical microstimulation (J. Lee et al., 2021). Carbon-based materials, particularly carbon nanotubes (CNTs) and graphene, have demonstrated optimal interaction and established a distinctive crosstalk with neuronal cells. There is an expectation that the use of carbon-based materials could render the interface less invasive (Li et al., 2023; Pampaloni et al., 2019).

Altogether, nanotech and micro-device technology have the potential to develop less invasive stimulation methods compared to traditional stimulation techniques such as ICMS. The ability to transduce various energy modalities into localized stimuli and interface with the neural system opens up new possibilities for modulating neural activity, even on a cell-type-specific basis. A class of neural stimulators, small enough for many to be implanted with relatively low risk to the patient, enables powerful therapeutic and neural-interface techniques (Piech et al., 2020). Neuroscience is ready for the systematic exploration of nanomaterials. But for this to happen, we need to break barriers between disciplines, with strong collaborations and interdisciplinary training for the next generations (Garcia-Etxarri and Yuste, 2021).

## 4.1 AI

Artificial Intelligence (AI) and neuroscience are developing hand in hand. The rapid progress in AI over the past decade has been remarkable, with undeniable societal impact and substantial repercussions on neuroscience. Experimental paradigms for BCI (e.g., cursor control, speech synthesis, neuroprosthetics of limbs/vision/audition/somatic sensation), have traditionally been developed and applied independently from those of AI. However, scientists now prefer to combine BCI and AI, enabling the efficient utilization of the brain's electric signals to control external devices (Macpherson et al., 2021; Zhang et al., 2020). In this section, we will introduce how AI contributes to brain/nerve stimulation.

The use of machine learning for reading from neural activity has increased significantly over the past decade and is proven, with "hand writing speller" being a good example, at 90 characters/min, currently considered as the most useful method for direct communicative output from the brain (Willett et al., 2021). While researchers persist in refining the performance of neural decoding models, the current results provide one starting point for future neural encoding work to "write in" neural information by patterned microstimulation (Heelan et al., 2019).

Despite the development and success of cochlear implants over several decades, there is reported wide inter-subject variability in speech perception. By combining the information theoretic methods with machine learning, Gao et al.'s model provided a flexible framework that can interrelate cochlear implant electrode discrimination with information transfer at the interface between the electrode array and auditory nerve fibers. This framework also enables the investigation of the extent to which key factors affect the performance of the cochlear implant model at the individual level (Gao et al., 2021). Baby et al. presented a hybrid approach where convolutional neural networks are combined with computational neuroscience to yield a real-time model for human cochlear mechanics, including level-dependent filter tuning (Baby et al., 2021). Their model accurately simulated human cochlear frequency selectivity and its dependence on sound intensity, an essential quality for robust speech intelligibility at negative speech-to-background-noise ratios. Crowson et al. reviewed machine learning use for cochlear implant and they found that applications of machine learning technologies involved speech/signal processing optimization, automated evoked potential measurement, postoperative performance/efficacy prediction and surgical anatomy location prediction (Crowson et al., 2020).

Prosthetic vision though retinal/cortex stimulation has far from ideal resolution. Recently, computer vision methods have contributed to optimizing retinal/cortex stimulation by extracting valuable features, including edge emphasis through object recognition and saliency detection, which is a kind of method or algorithm that mimics a human's ability to extract a region or object of interest (OOI) (Küçükoğlu et al., 2022; Wang et al., 2021).

Deep Brain Stimulation (DBS) is a well-established method that can enhance the quality of life for patients suffering from various neural disorders, including essential tremor, Parkinson's disease (PD), dystonia, Tourette syndrome, and obsessive-compulsive disorder. Machine learning has been successfully used for (1) DBS candidate selection; (2) programming optimization; (3) surgical targeting; and (4) insights into DBS mechanisms (Watts et al., 2020). Machine learning is also useful for adaptive close-loop DBS to optimize its stimulation from behavioral states and neural activity (Merk et al., 2022). There is a report which showed that fMRI together with their machine learning model can rapidly define the optimal DBS stimulation in PD patients (Boutet et al., 2021). Peralta et al. systematically reviewed the trend of use of machine learning for DBS (Peralta et al., 2021).

Google and Mayo clinic developed a new machine learning framework to probe how brain regions interact using single-pulse electrical stimulation that comprises the connectome (Miller et al., 2021). Personalized Bayesian optimization for transcranial alternating current stimulation (tACS) successfully searched, learned, and recommended parameters for effective neurointervention, as supported by behavioral, simulation, and neural data (van Bueren et al., 2021). Generative adversarial network (GAN) enhanced the accuracy of FUS through improved prediction of computed tomography (CT) from T1-weighted MRI (Koh et al., 2022; Liu et al., 2022).

For example, braille readers can reach 200-400 words/min, and the brain is adaptable enough to substitute modalities through training, such as learning sign language or reading musical scores. This means, for example, that if you study braille, you will be able to send language information via somatosensory brain area. AI, on the other hand, can convert signs into spoken language, translate images into natural language, and transform natural language into video and audio, and its development is progressing rapidly. With the advancement of writing technology for the peripheral nervous system and AI technology, the integration of AI technology and behavioral/cognitive training in a reciprocal combination could be the initial step towards developing a new method for converting information among modalities in the context of brain-to-brain communication.

AI is a powerful tool that is revolutionizing stimulation technology for the brain and nerves. Through enhanced signal analysis, optimization of stimulation parameters, and improved efficacy, AI is facilitating the development of more effective neural stimulation. As AI continues to advance, we can anticipate witnessing even more innovative and transformative applications of this technology in the field of neural stimulation for the brain and nerves.

**Conclusion and future directions**

Ultimately, the anticipation is that Brain-Computer-Interface (BCI) will usher in transformative capabilities, such as telepathy, facilitating the direct communication of intentions and notions, alongside human augmentation, exemplified by the transfer of memories into and out of the brain. However, the prospect of inscribing information onto the high hierarchical cortex remains beyond our current horizon.

Over the past two decades, ongoing research suggests that the initial advancements in BCI technology are likely to manifest through technological developments in peripheral nerves (sensory organs) or the lower cortex, where topographic maps such as retinotopy, tonotopy, or somatotopy are available. These developments are particularly relevant for individuals with physical disabilities. The foremost application under consideration for development is language transmission, with Intracortical Microstimulation (ICMS) garnering optimistic expectations. Potential targets include peripheral nerves (sensory organs) and the lower auditory cortex. In the short term, an increase in the information transfer rate (ITR) is anticipated, attributed to the proliferation of smaller input devices and a higher number of multiple channels. Optogenetics currently stands as the foremost research tool and is expected to find applications in human use in the mid- to long-term. Revolutionary changes are also anticipated through nanotechnology in the coming decade.

Analogous to the critical role of converting input signals in cochlear implants, the utilization of Artificial Intelligence (AI) for cortex stimulation holds promise for unveiling novel neurostimulation representations of external input. Recent achievements, such as the successful reading of letters from the motor cortex using an AI algorithm (Willett et al., 2021), suggest progress at a practically useful level. This accumulated knowledge may offer insights into the writing of information from the brain, fostering expectations for more interdisciplinary research between AI and neuroscience.

Fundamental innovations in neuroscience may still be needed to achieve nonverbal brain-to-brain connections in a way that goes beyond natural communication, through stimulation via the higher cortex, similar to what is required for visual neuro-prosthetics.

**Acknowledgements**

This work was supported by JST, Moonshot R&D Grant Number JPMJMS2012.